\documentclass[11pt]{article}
\usepackage{amsmath}
\input amssym

\begin{document}
\def\prg#1{\medskip\noindent{\bf #1}}     \def\ra{\rightarrow}
\def\lra{\leftrightarrow}        \def\Ra{\Rightarrow}
\def\nin{\noindent}              \def\pd{\partial}
\def\dis{\displaystyle}          \def\inn{\,\rfloor\,}
\def\Lra{{\Leftrightarrow}}
\def\Leff{\hbox{$\mit\L_{\hspace{.6pt}\rm eff}\,$}}
\def\bull{\raise.25ex\hbox{\vrule height.8ex width.8ex}}
\def\ads{(A)dS}

\def\G{\Gamma}        \def\S{\Sigma}        \def\L{{\mit\Lambda}}
\def\D{\Delta}        \def\Th{\Theta}
\def\a{\alpha}        \def\b{\beta}         \def\g{\gamma}
\def\d{\delta}        \def\m{\mu}           \def\n{\nu}
\def\th{\theta}       \def\k{\kappa}        \def\l{\lambda}
\def\vphi{\varphi}    \def\ve{\varepsilon}  \def\p{\pi}
\def\r{\rho}          \def\Om{\Omega}       \def\om{\omega}
\def\s{\sigma}        \def\t{\tau}          \def\eps{\epsilon}
\def\nab{\nabla}      \def\btz{{\rm BTZ}}   \def\heps{\hat\eps}

\def\tG{{\tilde G}}   \def\tR{{\tilde R}}
\def\bA{{\bar A}}     \def\bF{{\bar F}}     \def\bG{{\bar G}}
\def\bt{{\bar\tau}}   \def\bg{{\bar g}}     \def\cT{{\cal T}}
\def\cL{{\cal L}}     \def\cM{{\cal M }}    \def\cE{{\cal E}}
\def\cH{{\cal H}}     \def\bcH{{\bar\cH}}   \def\hcH{\hat{\cH}}
\def\cK{{\cal K}}     \def\hcK{\hat{\cK}}   \def\tcL{{\tilde\cL}}
\def\cO{{\cal O}}     \def\hcO{\hat{\cal O}}
\def\cV{{\cal V}}     \def\bV{{\bar V}}
\def\cB{{\cal B}}     \def\heps{\hat\epsilon}

\def\bb{{\bar b}}     \def\bt{{\bar\t}}     \def\br{{\bar\r}}
\def\bpi{{\bar\pi}}   \def\bom{{\bar\om}}   \def\bphi{{\bar\phi}}
\def\bJ{{\bar J}}     \def\bl{{\bar\l}}     \def\tom{{\tilde\omega}}

\def\hR{{\hat R}}     \def\hpi{{\hat\pi}}   \def\hPi{{\hat\Pi}}

\def\bi{{\bar\imath}} \def\bj{{\bar\jmath}} \def \bk{{\bar k}}
\def\bm{{\bar m}}     \def\bn{{\bar n}}     \def\bl{{\bar l}}
\def\bT{{\overline{T\mathstrut}}}
\def\bR{{\overline{R\mathstrut}}}
\def\cR{{\cal R}}       \def\cA{{\cal A}}
\def\hA {{\hat A}}      \def\hB{{\hat B}}   \def\hR{{\hat R}{}}
\def\hOm{{\hat\Omega}}  \def\hL{{\hat\L}}

\def\tb{{\tilde b}}    \def\tA{{\tilde A}}  \def\tV{{\tilde V}}
\def\tT{{\tilde T}}    \def\tR{{\tilde R}}  \def\tv{{\tilde V}}
\def\tcA{{\tilde\cA}}  \def\ts{\tilde\s}    \def\tcV{\tilde\cV}

\def\irr#1#2{\hspace{0.5pt}{}^{#1}\hspace{-1.6pt}#2}
\def\mb#1{\hbox{{\boldmath $#1$}}}
\def\fT{\mb{T}}       \def\fR{\mb{R}}       \def\fV{\mb{V}}
\def\fA{\mb{\cA}}     \def\tot{{\rm tot}}   \def\can{{\rm c}}
\def\vsm{\vspace{-8pt}}

\def\nb{\marginpar{\bf\Huge ?}}
\def\nn{\nonumber}
\def\be{\begin{equation}}             \def\ee{\end{equation}}
\def\ba#1{\begin{array}{#1}}          \def\ea{\end{array}}
\def\bea{\begin{eqnarray} }           \def\eea{\end{eqnarray} }
\def\beann{\begin{eqnarray*} }        \def\eeann{\end{eqnarray*} }
\def\beal{\begin{eqalign}}            \def\eeal{\end{eqalign}}
\def\lab#1{\label{eq:#1}}             \def\eq#1{(\ref{eq:#1})}
\def\bsubeq{\begin{subequations}}     \def\esubeq{\end{subequations}}
\def\bitem{\begin{itemize}}           \def\eitem{\end{itemize}}

\title{Poincar\'e gauge theory in 3D:
       canonical stability of the scalar sector}

\author{M. Blagojevi\'c and B. Cvetkovi\'c\,\footnote{Based on a talk by MB at
\emph{New ideas for unsolved problems II}, Div\v cibare,
22--24 Sep 2013, Serbia.}\,
\footnote{Email addresses: {\tt mb@ipb.ac.rs, cbranislav@ipb.ac.rs}}\\
University of Belgrade, Institute of Physics, P. O. Box 57,
11001 Belgrade, Serbia}
\date{\today}
\maketitle

\begin{abstract}

We outline the results of the canonical analysis of the
three-dimensio\-nal Poincar\'e gauge theory, defined by the general
parity-invariant Lagrangian with eight free parameters \cite{x11}. In the
scalar sector, containing scalar or pseudoscalar (A)dS modes, the
stability of the canonical structure under linearization is used to
identify dynamically acceptable values of the parameters.
\end{abstract}

\section{Introduction}
\setcounter{equation}{0}

Models of three-dimensional (3D) gravity, pioneered by Staruskiewicz
\cite{x1}, were introduced to help us in clarifying highly complex
dynamical behavior of the realistic four-dimensional general relativity
(GR). In the last three decades, they led to a number of outstanding
results \cite{x2}. However, in the early 1990s, Mielke and Baekler
\cite{x3} proposed a new, non-Riemannian approach to 3D gravity, based on
the Poincar\'e gauge theory (PGT) \cite{x4}. In PGT, the basic
gravitational variables are the triad $b^i$ and the Lorentz connection
$A^{ij}$ (1-forms), and their field strengths are the torsion
$T^i:=db^i+A^i{_j}b^j$ and the curvature $R^{ij}:=dA^{ij}+A^i{_m}A^{mj}$
(we omit the exterior product sign for simplicity). In contrast to the
traditional GR, with an underlying Riemannian geometry of spacetime, the
PGT approach is characterized by a Riemann--Cartan geometry, with both the
curvature and the torsion of spacetime as carriers of the gravitational
dynamics. Thus, PGT allows exploring the interplay between gravity and
geometry in a more general setting.

Three-dimensional GR with or without a cosmological constant, as well as
the Mielke--Baekler (MB) model, are \emph{topological} theories without
propagating modes. From the physical point of view, such a degenerate
situation is certainly not quite realistic. Including the
\emph{propagating modes} in PGT is achieved quite naturally by using
Lagrangians quadratic in the field strengths \cite{x5,x6}.

Since the general parity-invariant PGT Lagrangian in 3D is defined by
eight free para\-me\-ters \cite{x6}, it is a theoretical challenge to find
out which values of the parameters are allowed in a viable theory. The
simplest approach to this problem is based on the \emph{weak-field
approximation} around the Minkowski background \cite{x5}. However, one
should be very careful with the interpretation of these results, since the
weak-field approximation does not always lead to a correct identification
of the physical degrees of freedom.

The constrained Hamiltonian method \cite{x7,x4} is best suited for
analyzing dynamical content of gauge theories of gravity, respecting fully
their \emph{nonlinear structure}. However, as noticed by Yo and Nester
\cite{x8,x9}, it may happen, for some ranges of parameters, that the
canonical structure of a theory (the number and/or type of constraints) is
changed after linearization in a way that affects its physical content,
such as the number of physical degrees of freedom. Such an effect is
called the phenomenon of \emph{constraint bifurcation}. Based on the
\emph{canonical stability under linearization} as a criterion for an
acceptable choice of parameters, Shie et al. \cite{x10} proposed a PGT
cosmological model that offers a convincing explanation of dark energy as
an effect induced by torsion.

In this note, we use the constrained Hamiltonian formalism to study (a)
the phenomenon of constraint bifurcation and (b) the stability under
linearization of the general parity-invariant PGT in 3D, in order to find
out the parameter values that define consistent models of 3D gravity with
propagating torsion. Because of the complexity of the problem, we restrict
our attention to the scalar sector, with $J^P=0^+$ or $0^-$ modes, defined
with respect to the (A)dS background \cite{x11}.

The following conventions are of particular importance for our canonical
analysis. Let $\cM$ be a 3D manifold (spacetime) with local coordinates
$x^\m=(x^0,x^\a)$, and $h_i=h_i{^\m}\pd_\m$ a Lorentz frame on it. Then,
if $\S$ is a 2D spacelike surface with a unit normal $n_k$, each tangent
vector $V_k$ of $\cM$ can be decomposed in terms of its normal and
parallel component with respect to $\S$:
$$
V_k=n_kV_\perp+V_\bk\, ,\quad {\rm where}\quad
    V_\perp:=n^mV_m\, ,\quad V_\bk=h_k{^\a}V_\a\, .
$$
Note that $V_\bk$ does not contain the time component of $V_\m$.

\section{Quadratic PGT and its scalar modes}
\setcounter{equation}{0}

Assuming parity invariance, the dynamics of 3D gravity
with propagating torsion is determined by the gravitational Lagrangian
\bsubeq\lab{1}
\be
L_G=-a\ve_{ijk}b^iR^{jk}
    -\frac{1}{3}\L_0\ve_{ijk}b^ib^jb^k+L_{T^2}+L_{R^2}\, ,      \lab{1a}
\ee
where $a=1/16\pi G $, $\L_0$ is a free parameter (bare cosmological
constant), the pieces quadratic in the field strengths read
\bea
L_{T^2}&:=&T^i{}^*\left(a_1{}^{(1)}T_i
           +a_2{}^{(2)}T_i+a_3{}^{(3)}T_i\right)\, ,            \nn\\
L_{R^2}&:=&\frac{1}{2}R^{ij}\left( b_4{}^{(4)}R_{ij}
        +b_5{}^{(5)}R_{ij}+b_6{}^{(6)}R_{ij}\right)\, ,         \lab{1b}
\eea
\esubeq
and $^{(n)}T_i$ and $^{(n)}R_{ij}$ are irreducible components of $T^i$ and
$R^{ij}$ \cite{x6}. Being interested only in the gravitational degrees of
freedom, we disregard the matter contribution.

Particle spectrum of the theory around the Minkowski background $M_3$ is
already known \cite{x5,x6}. Restricting our attention to the scalar
sector, we display here the masses of the spin-$0^+$ and $0^-$ modes:
\bsubeq\
\be
m_{0^+}^2=\frac{3a(a+a_2)}{a_2(b_4+2b_6)}\, ,\qquad
m_{0^-}^2=\frac{3a(a+2a_3)}{(a_1+2a_3)b_5}\, .                  \lab{2a}
\ee
These modes have finite masses and propagate if
\be
a_2(b_4+2b_6)\ne 0\, ,\qquad  (a_1+2a_3)b_5\ne 0\, ,            \lab{2b}
\ee
\esubeq
respectively.

Transition to the (A)dS background is straightforward; it generalizes the
mass formulas \eq{2a} by introducing a dependence on the parameter $q$
that measures the strength of the background curvature \cite{x11}, but the
propagation conditions for the scalar modes remain the \emph{same} as in
\eq{2b}. As we shall see in the next section, the conditions \eq{2b},
derived in the \emph{weak-field approximation}, have a critical role also
in the canonical analysis of the \emph{full nonlinear theory}.

\section{Primary if-constraints}

The canonical momenta corresponding to the basic dynamical variables
$(b^i{_\mu},A^{ij}{}_\m)$ are defined by $\pi_i{^\m}:={\pd\tcL}/{\pd(\pd_0
b^i{_\m})}$ and $\Pi_{ij}{^\m}:={\pd\tcL}/{\pd(\pd_0 A^{ij}{}_\m)}$,
respectively. Since the torsion and the curvature do not involve the
velocities $\pd_0b^i{_0}$ and $\pd_0 A^{ij}{_0}$, one obtains the primary
constraints
\be
\pi_i{^0}\approx 0\, ,\qquad \Pi_{ij}{^0}\approx 0\, ,
\ee
which are always present, independently of the values of coupling
constants (``sure" constraints). If the Lagrangian \eq{1} is singular with
respect to some of the remaining velocities $\pd_0 b^i{_\a}$ and $\pd_0
A^{ij}{_\a}$, one obtains further primary constraints, known as the
primary ``if-constraints" (ICs).

The gravitational Lagrangian \eq{1} depends on the time derivative $\pd_0
b^i{_\a}$ only through the torsion tensor, appearing in $L_{T^2}$. The
system of equations defining the parallel gravitational momenta
$\hpi_i{^\bk}=\pi_i{^\a}b^k{_\a}$ ($\hpi_i{^\bk}n_k=0$) can be decomposed
into irreducible parts with respect to the group of two-dimensional
spatial rotations in $\S$:
\bsubeq\lab{4}
\bea
&&\phi_{\perp\bk}:=\frac{\hpi_{\perp\bk}}J-(a_2-a_1)T^\bm{}_{\bm\bk}
    =(a_1+a_2)T_{\perp\perp\bk} \,,                             \lab{4a}\\
&&\irr{S}{\phi}:=\frac{{}^S\hpi}J=-2a_2T^\bm{}_{\bm\perp}\,,    \\
&&\irr{A}{\phi}_{\bi\bk}:=\frac{{}^A\hpi_{\bi\bk}}J
  -\frac{2}{3}(a_1-a_3)T_{\perp\bi\bk}=
  -\frac{2}{3}(a_1+2a_3)T_{[\bi\bk]\perp}\,,                    \\
&&\irr{T}{\phi}_{\bi\bk}:=\frac{{}^T\hpi_{\bi\bk}}J
  = -2a_1\irr{T}{T}_{\bi\bk\perp}\, ,
\eea
\esubeq
where the terms depending on the velocities $\pd_0 b^i{_\a}$ are moved to
the right-hand sides. If the critical parameter combinations appearing on
the right-hand sides of Eqs. \eq{4} vanish, the corresponding expressions
$\phi_K$ become additional primary constraints.

Similar analysis can be applied to the equations defining the parallel
gravitational momenta $\hPi_{ij}{^\bk}=:\Pi_{ij}{^\a}b^k{_\a}$
($\hPi_{ij}{^\bk}n_k=0$), leading to an additional set of primary
constraints $\Phi_K$. The complete set of primary ICs, including their
spin-parity characteristics ($J^P$), is shown in Table 1.
\begin{center}
\doublerulesep 1.6pt
\begin{tabular}{l l l}
\multicolumn{3}{c}{Table 1. Primary if-constraints}        \\
\hline\hline\rule[-5.5pt]{0pt}{20pt}
Critical conditions & Primary constraints &$J^P$           \\
\hline\rule[-7pt]{0pt}{21pt}
$a_2=0$        &$\irr{S}{\phi}\approx 0$ &                 \\[-1.2ex]
~$b_4+2b_6=0$  &$\irr{S}{\Phi}_{\perp}\approx 0$
               & \raisebox{1.6ex}{$0^+$}                   \\
\hline\rule[-7pt]{0pt}{21pt}
$a_1+2a_3=0$   &$\irr{A}{\phi}_{\bi\bk}\approx 0$ &        \\[-1.2ex]
~$b_5=0$       &$\irr{A}{\Phi}_{\perp\bi\bk}\approx 0$
               & \raisebox{1.6ex}{$0^-$}                   \\
\hline\rule[-7pt]{0pt}{21pt}
$a_1+a_2=0$  &${\phi}_{\perp\bk}\approx 0$ &               \\[-1.2ex]
~$b_4+b_5=0$ &$\irr{V}{\Phi}_\bk\approx 0$
             & \raisebox{1.6ex}{$1$}                       \\
\hline\rule[-7pt]{0pt}{21pt}
$a_1=0$  &$\irr{T}{\phi}_{\bi\bk}\approx 0$ &              \\[-1.2ex]
~$b_4=0$ &$\irr{T}{\Phi}_{\perp\bi\bk}\approx 0$
         & \raisebox{1.6ex}{$2$}                           \\
\hline\hline
\end{tabular}
\end{center}
This classification has a remarkable interpretation: whenever a pair of
the ICs with specific $J^P$ is absent, the corresponding dynamical mode is
liberated to become a \emph{physical degree of freedom} (DoF). Thus, for
$a_2(b_4+2b_6)\ne 0$, the spin-$0^+$ ICs are absent, and the spin-$0^+$
mode becomes physical. Similarly, $(a_1+2a_34)b_5\ne 0$ implies that the
spin-$0^-$ mode is physical. These results, referring to the full
nonlinear theory, should be compared to \eq{2b}.

\prg{Remark.} Once we know the complete set of primary ICs, we can
apply Dirac's consistency algorithm to obtain the secondary constraints,
and so on.

\section{Spin-\mb{0^+} sector}

As one can see from Table 1, the spin-$0^+$ degree of freedom propagates
for $a_2(b_4+2b_6)\ne 0$. In order to investigate dynamical features of
this sector, we adopt somewhat simplified conditions:
\bsubeq
\be
a_2,b_6\ne 0\, ,\qquad a_1=a_3=b_4=b_5=0\, .                    \lab{5a}
\ee
While such a ``minimal" choice simplifies the calculations, it is not
expected to influence any essential aspect of the spin-$0^+$ dynamics
\cite{x8,x9}.

\subsection*{Generic case}

Now, we turn to the canonical analysis. First, the form of the Hamiltonian
implies that the kinetic energy density is positive definite (no
``ghosts") if
\be
a_2>0\,,\qquad b_6>0\, .                                        \lab{5b}
\ee
\esubeq
Second, in the simple, \emph{generic} situation, when all of the ICs are
second class (their number is $N_2=10$), the complete set of constraints
is given in Table 2.
\begin{center}
\doublerulesep 1.6pt
\begin{tabular}{lll}
\multicolumn{3}{c}{Table 2. Generic constraints
                                 in the $0^+$ sector} \\
                                                      \hline\hline
\rule{0pt}{12pt}
& First class \phantom{x} & ~~Second class\phantom{x} \\
                                                      \hline
\rule[-1pt]{0pt}{16pt}
Primary
  ~& $\pi_i{^0}$, $\Pi_{ij}{^0}$
   & ~~$\irr{V}{\Phi}_\bi;
        \irr{A}{\phi},\irr{A}{\Phi},\irr{T}{\phi},\irr{T}{\Phi}$  \\
                                                      \hline
\rule[-1pt]{0pt}{19pt}
Secondary\phantom{x}
  ~& $\cH'_{\perp}$, $\cH'_\a$, $\cH'_{ij}$ & ~~$\chi_\bi$\\
                                                      \hline\hline
\end{tabular}
\end{center}
As always, the Hamiltonian constrains  $\cH'_{\perp}$, $\cH'_\a$ and
$\cH'_{ij}$ are first class. With $N=2\times 9$ field components,
$N_1=2\times 6$ first class constraints and $N_2=10$ second class
constraints, the dimension of the phase space is $N^*=2N-2N_1-N_2=2$, and
the theory exhibits a single Lagrangian DoF.

\subsection*{Constraint bifurcation}

To clarify the term ``generic" used above, we calculate the determinant of
the $10\times 10$ matrix $\D^+_{MN}=\{X'_M,X'_N\}$, where $X_M'$ is the
set of all ICs shown in Table 2. The result is
\be
\D^+\sim W^{10}\left(W-a_2\right)^4 \quad{\rm where}\quad
         W:=\frac{\irr{S}{\Pi}_\perp}{4J}\, .                   \lab{6}
\ee
The generic situation corresponds to $\D^+\ne 0$. However, the determinant
$\D^+$, being a field-dependent object, may vanish in some regions of
spacetime, changing thereby the number and/or type of constraints and the
number of physical DoF, as compared to the situation described in Table 2.
This phenomenon of \emph{constraint bifurcation} can be fully understood
by analyzing dynamical behavior of the critical factors $W$ and $W-a_2$,
appearing in $\D^+$.

Assuming that $W$ is an analytic function globally, on the whole spacetime
manifold $\cM$, the analysis of the field equations
\be
-(W-a_2)V_k +2\pd_k(W-a_2)\approx 0\, ,                         \lab{7}
\ee
leads to the following conclusion \cite{x11}:
\bitem
\item[\bull] If there is a point in $\cM$ at which $W-a_2\ne 0$, then
$W-a_2\ne 0$ globally.
\eitem
Hence, by choosing the initial data so that $W-a_2\ne 0$ at $x^0=0$, it
follows that $W-a_2$ stays nonvanishing for any $x^0>0$. The surface
$W-a_2=\frac{1}{6}b_6R-a-a_2\approx 0$ (on shell) is a dynamical barrier
that the spin-$0^+$ field cannot cross. Moreover, since $a_2$ is positive,
see \eq{5b}, we have:
\bitem
\item[\bull] By choosing $W-a_2>0$ at $x^0=0$, it follows that $W\ne 0$
globally.
\eitem
Thus, with a suitable choice of the initial data, one can ensure the
generic condition $\D^+\ne 0$ to hold \emph{globally}, whereupon the
constraint structure is described exactly as in Table 2. Any other
situation, with $W=0$~ or ~$W-a_2=0$, would not be acceptable---it would
have a variable constraint structure over the spacetime, the property that
could not survive the process of linearization.

\subsection*{Stability under linearization}

Now, we compare the canonical structure of the full nonlinear theory with
its weak-field approximation around maximally symmetric background. With
the background values $\bar R=-6q$ and $\bar W= \frac{1}{6}b_6\bar R-a$,
the lowest-order critical factors take the form
$$
\bar W=-(a+qb_6)\, ,\qquad \bar W-a_2=-(a+a_2+qb_6)\, ,
$$
which leads to the results shown in Table 3 \cite{x11}.
\begin{center}
\doublerulesep 1.6pt
\begin{tabular}{ccccl}
\multicolumn{5}{c}{
               Table 3. Canonical stability in the $0^+$ sector}\\
                                                      \hline\hline
\rule{-1pt}{16pt}
    & $a+qb_6$ &~$a+a_2+qb_6$ & DoF & stability       \\[2pt]
                                                      \hline
\rule[-1pt]{0pt}{16pt}
(a) & $\ne 0$  & $\ne 0$      & 1 & stable            \\[2pt]
                                                      \hline
\rule[-1pt]{0pt}{16pt}
(b) & $=0$     & $\ne 0$      & 0 & unstable          \\[2pt]
                                                      \hline
\rule[-1pt]{0pt}{16pt}
(c) & $\ne 0$  & $=0$         & 1 & stable*           \\[2pt]
                                                      \hline\hline
\end{tabular}
\end{center}
Based on the conditions \eq{5a}, the spin-$0^+$ mass formula for $q\ne 0$
takes the form:
$$
m_{0^+}^2=\frac{3(a-qb_6)(a+a_2+qb_6)}{2a_2b_6}\, .
$$
Now, a few comments are in order: (a) the nature of constraints remains
the same as in Table 2, which implies the stability under linearization;
(b) all if-constraints become first class, but only 6 of them remain
independent, which leads to $N^*=0$ (instability); (c) the massless
nonlinear theory, defined by the condition $a+a_2+qb_6=0$, is essentially
stable under linearization.

\section{Concluding remarks}
\setcounter{equation}{0}

\noindent
--- By investigating fully nonlinear "constraint bifurcation" effects, as
well as the canonical stability under linearization, we were able to
identify the set of dynamically acceptable values of parameters for the
spin-$0^+$ sector of PGT, as shown in Table 3.

\noindent
--- On the other hand, the spin-$0^-$ sector is canonically
unstable for any choice of parameters; for more details, see Ref.
\cite{x11}.

\noindent
--- Further analysis of higher spin modes is left for future studies.

\section*{Acknowledgements}

We thank Vladimir Dragovi\'c for a helpful discussion This work was
supported by the Serbian Science Foundation under Grant No. 171031.


\end{document}